\documentclass[12pt]{article}
\usepackage{amsmath,amssymb,amsthm}
\usepackage{mathtools}
\usepackage{graphics,epsfig}
\usepackage{hyperref}
\usepackage{natbib}
\usepackage{color}
\usepackage{graphicx}
\usepackage{caption}
\usepackage{subcaption}
\usepackage{float}
\usepackage{mathrsfs}
\usepackage{multirow}
\usepackage{comment}
\usepackage{setspace}
\usepackage{bbm}
\usepackage{bm}
\usepackage{amsfonts}
\usepackage{xcolor}
\usepackage{pslatex}
\usepackage{setspace}
\usepackage{bbm}
\def \E {\mathbb{E}}

\def \a {\alpha}
\def \be {\beta}


\begin{document}

  \title{\bfseries Stochastic modeling of cyclic cancer treatments under common noise }

  \author{
Jason Sonith\footnote{Corresponding author, {\small\texttt{jis2123@jagmail.southalabama.edu}}}\; \footnote{Department of Computer Science,  University of South Alabama, Shelby Hall, Suite 2101, 150 Student Services Drive, Mobile, AL, 36688,
United States.}
}

\date{\today}
\maketitle

\begin{abstract}
Path integral control is an effective method in cancer drug treatment, providing a structured approach to handle the complexities and unpredictability of tumor behavior. Utilizing mathematical principles from physics, this technique optimizes drug delivery in environments influenced by randomness. It takes into account the intricate interactions between cancer cells, healthy tissues, and the immune system, as well as factors such as patient-specific characteristics and tumor diversity. Path integral control offers tailored solutions to these issues, enabling the design of drug dosing regimens that enhance therapeutic effectiveness while minimizing side effects. Its flexibility makes it a valuable tool in creating personalized, precision-driven therapies, ultimately improving patient outcomes in cancer treatment. In this paper we give a review about the current status of path integral control in cancer research.
\end{abstract}

\section*{Introduction:}
Mathematical models of population growth based on nonlinear ordinary differential equations have been extensively explored due to their ability to capture the fundamental aspects of complex biological interactions despite their relative simplicity. These models often provide insights into proliferation phenomena \citep{ochab2004population}. However, biological systems are rarely purely deterministic; natural systems are influenced by various types of stochastic fluctuations \citep{esmaili2024optimal}. Broadly, these fluctuations can be classified into two categories: first, internal noises, such as thermodynamic fluctuations, which reflect variability in system parameters at thermodynamic equilibrium, and second, external noises, often conformational in nature, which arise from environmental time-dependent changes \citep{ochab2004population}.

To study the dynamics of biological systems undergoing conformational transitions, a \emph{mesoscopic} approach using stochastic differential equations is often employed. This framework describes system kinetics analogously to the \emph{Kramers} scenario, where a Brownian particle moves through an effective free energy landscape. In many instances, this potential landscape is not static but is influenced by random fluctuations with a characteristic time scale comparable to that of the particle’s traversal across energy barriers between stable states \citep{doering1992resonant,bulsara1996cooperative,astumian2002protein,iwaniszewski2003mean}. Examples of such dynamics include the escape of oxygen or carbon monoxide from myoglobin following dissociation, ion channel behavior in lipid membranes, and protein folding \citep{astumian2002protein}. Another example is the interaction between external noise and a collective variable in the context of immune system-mediated cancer surveillance \citep{garay1978kinetic}.

Most tumor cells express antigens recognized as foreign by the immune system, which can trigger responses mediated by immune cells like T-lymphocytes, as well as other cells, such as macrophages or natural killer cells. Tumor damage occurs as these specialized cells infiltrate the tumor, initiating cytotoxic activity against cancer cells \citep{esmaili2017optimal}. The sequence of cytotoxic reactions between immune cells and tumor tissue can be approximated by a saturating, enzyme-like process, with time evolution similar to Michaelis-Menten kinetics \citep{pramanik2024dependence}. Importantly, variability in the kinetic parameters governing this process significantly influences tumor eradication outcomes \citep{garay1978kinetic,ochab2004population}.

Stochastic control is essential in cancer research as it addresses the unpredictability and variability inherent in tumor progression and treatment outcomes \citep{pramanik2024parametric}. Factors such as genetic mutations, epigenetic shifts, the tumor microenvironment, and individual immune responses introduce significant randomness, making traditional deterministic models inadequate for capturing the full range of possible scenarios \citep{zhao2010parabolic}. By leveraging stochastic control methods, scientists can integrate these uncertainties into predictive frameworks, enabling simulations of diverse outcomes and optimization of therapeutic strategies in real-time \citep{pramanik2021optimization,pramanik2021optimal,pramanik2023scoring}. For example, in treatments like chemotherapy or radiotherapy, stochastic models can identify optimal dosing regimens that effectively reduce tumors while minimizing harmful side effects, even amidst varying patient responses \citep{pramanik2020motivation}. These approaches also support the development of adaptive therapies, which adjust treatment dynamically based on a patient’s evolving condition, ensuring a more personalized and effective approach. Furthermore, representing tumor dynamics and treatment effects as stochastic processes allows researchers to explore critical phenomena, such as the emergence of resistance and the spread of metastases. This understanding aids in crafting strategies that not only address current cancer cells but also anticipate future challenges \citep{pramanik2020optimization,pramanik2024optimization}. Overall, applying stochastic control in cancer research drives innovation in therapy design, enhances patient outcomes, and advances the pursuit of precision medicine by tailoring interventions to the unique characteristics of each individual case.

McKean-Vlasov dynamics \cite{mckean1966class} plays a vital role in stochastic control, especially within the framework of mean field games (MFGs) \citep{lasry2007mean}, as it provides a powerful method for modeling the probabilistic interactions of a large number of agents. In systems involving numerous entities, such as those found in financial markets, social networks, or large-scale energy infrastructures, it becomes impractical to explicitly model the interactions between every pair of agents \citep{polansky2021motif,pramanik2024estimation}. McKean-Vlasov dynamics overcomes this limitation by focusing on the evolution of a representative agent’s state, which is influenced by the overall statistical distribution of the population rather than direct, individual interactions \citep{hua2019assessing}. This enables the modeling of collective behavior in a way that remains computationally efficient. Additionally, the dynamics naturally account for feedback loops, where each agent’s actions shape the population distribution, which then affects the environment experienced by all agents. In the context of MFGs, this feedback mechanism is essential for capturing equilibrium states, as agents optimize their personal objectives while simultaneously considering the cumulative effect of their choices on the broader system \citep{carmona2013control,carmona2015forward}.

Furthermore, McKean-Vlasov dynamics offers a robust mathematical framework to address intricate dependencies in control problems, particularly when the costs or dynamics of an individual agent are influenced by mean-field factors such as the average state or overall distribution of the population \citep{pramanik2022lock,pramanik2022stochastic}. This capability is especially critical in scenarios where strategic decisions rely not only on an agent’s own state but also on expectations about the collective behavior of the system. For example, in finance or economics, agents may adjust investment strategies based on average market trends, while in energy systems, consumption strategies might be optimized in response to aggregate demand patterns \citep{pramanik2024semicooperation,pramanik2023path1,pramanik2023path}. Additionally, McKean-Vlasov dynamics facilitates modeling under conditions of uncertainty, allowing for the creation of policies that remain effective in the face of stochastic variations and mean-field interactions. Importantly, this approach bridges the gap between microscopic (individual agent) and macroscopic (population-wide) perspectives, linking personal decision-making with emergent system-level behaviors. This dual nature makes McKean-Vlasov dynamics an essential tool for developing both theoretical understanding and practical applications of stochastic control in mean field contexts, solidifying its role in modern system analysis.

McKean-Vlasov dynamics with common noise represents a significant advancement in stochastic control theory because it incorporates shared random influences affecting an entire population of interacting agents, thereby accounting for global uncertainties that impact all participants collectively \citep{pramanik2016tail,pramanik2021effects}. In practical systems such as financial markets, environmental frameworks, or large-scale networks, agents are influenced not only by their individual stochastic variations and the overall population behavior but also by external random events like macroeconomic shifts, regulatory changes, or environmental fluctuations. Integrating common noise into McKean-Vlasov dynamics allows for the modeling of these correlated uncertainties and their effects on collective decision-making and behavior. This is particularly relevant in mean field games, where common noise plays a crucial role in shaping equilibrium states and influencing the strategies agents employ. For instance, in investment management, a systemic shock such as a sudden interest rate adjustment affects all market participants, leading to correlated changes in their strategies. Similarly, in energy systems, variations in renewable energy production due to weather conditions introduce shared uncertainties that impact the stability of the grid and consumption patterns across agents \citep{pramanik2024motivation}. Furthermore, common noise facilitates the analysis of how uncertainties propagate and the assessment of systemic risks, both of which are essential for evaluating the resilience of interconnected systems. By distinguishing between idiosyncratic noise, which affects individual agents, and shared noise, which drives system-wide dynamics, the framework adds realism to models \citep{kaznatcheev2017cancer}. This dual perspective is vital for developing robust control strategies capable of addressing both individual-level variability and systemic risks, making McKean-Vlasov dynamics with common noise a foundational element in the progression of stochastic control theory and its practical implementations.

\section*{An overview of the problem:}
Consider a fixed and finite time \( t > 0 \), representing the duration of a cancer treatment. In the context of a measurable space \((\Omega, \mathcal{F})\), let \(\mathcal{P}(\Omega, \mathcal{F})\) denote the probability measure defined on \((\Omega, \mathcal{F})\), where \(\Omega\) represents the sample space corresponding to the cancer's state, and \(\mathcal{F}\) is the Borel \(\sigma\)-algebra, \(\mathcal{B}(\Omega)\) \citep{carmona2016mean}. Define \(\mathfrak{C}^n = C([0,t] \mid \mathbb{R}^n)\) as the set of continuous functions mapping \([0,t]\) into \(\mathbb{R}^n\). Moreover, by denoting \(Lip(\mathbb R^n)\) the space of Lipschitz real valued function and by $||.||_{Lip}$ the corresponding norm 
\[
||b||_{Lip}:=\sup_{\substack{x,\tilde x\in\mathcal X\\ x\neq\tilde x}}\frac{|b(x)-b(\tilde x)|}{|x-\tilde x|},
\]
where $x$ is a value a value of the cancer state $X\subset\mathcal X$ which takes the value from $\mathbb R^n$. The Borel probability measure on $\mathbb R^n$ with the finite first moment  by $\tilde {\mathcal P}(\mathbb R^n)$ and distance is measured by first order Wasserstein distance defined by
\[
\mathcal V_1(\tau,\hat\tau)=\inf_{\zeta\in\Pi(\tau,\hat\tau)}\int_{\mathbb R^n\times\mathbb R^n}|x-\tilde x|\ \zeta(dxd\tilde x),
\]
where $\Pi(\tau,\hat\tau)$ is the set of couplings of $\tau$ and $\hat\tau$. In other words, it is a set of probability measures on $\mathbb R^n\times\mathbb R^n$ whose first marginal distribution is $\tau$ and the second marginal is $\hat\tau$ \citep{conforti2023coupling}. Now consider $\left(\Omega,\{\mathcal F_s\}_{s\in[0,t]},\mathcal F,\mathcal P\right)$ filtration probability space and $\{B(s)\}_{s\in[0,t]}$ a standard $n$-dimensional and $\mathcal F_s$-adapted Brownian motion. Given $0\leq s\leq t$, an $\mathbb R^n$ process is denoted as $\{u(\nu)\}_{\nu\in[s,t]}$, and furthermore, it is indeed an admissible cancer treatment if $\{u(\nu)\}_{\nu\in[s,t]}$ is progressively measurable such that
\[
\E\left\{\int_s^t|u(\nu)|^\rho d\nu\right\}<\infty,\ \ \forall\rho\in\mathbb N.
\]
Moreover, we denote $\mathcal U_[s,t]$ the set of admissible treatments. Our main objective is to solve
\begin{equation}\label{1}
 \inf_u\mathcal J (x_0,u):=\inf_u\ \E\left\{\int_0^tm[s, X(s),u(s),\tau(s)]ds+\tilde m[X(t),\tau(t)]\bigg | \mathcal F_0\right\},   
\end{equation}
subject to a McKean-Vlasov dynamics represented by the SDE
\begin{equation}\label{2}
 dX(s)=\mu[s, X(s),u(s),\tau(s)]ds+\sigma[s, X(s),u(s),\tau(s)]d B(s)+\sigma_0[s, X(s),u(s),\tau(s)]d W(s),  
\end{equation}
where $W(s)$ is a Wiener process called the common noise. For Equation \eqref{2} assume the following mappings
\begin{align}\label{3}
 & (\mu, m):[0,t]\times\mathbb R^n\times\mathcal U\times\mathcal P(\mathbb R^n) \mapsto\mathbb R^n\times\mathbb R,\notag\\
 &(\bm\sigma,\bm\sigma_0):[0,t]\times\mathbb R^n\times\mathcal U\times\mathcal P(\mathbb R^n) \mapsto\mathbb R^n\times\mathbb R,\notag\\
 &\tilde m:\mathbb R^n\times\mathcal P(\mathbb R^n)\mapsto\mathbb R.
\end{align}
In Equation \eqref{2}, $\mu$, $\bm\sigma$ and $\bm\sigma_0$ represent the drift, state specific diffusion and common diffusion components of the system, respectively. The variable $u \in \mathcal{U}$ denotes an adaptive treatment, while $m$ is the treatment cost function with a terminal value $\tilde{m}$ that corresponds to two possible outcomes: treatment success or failure. $\tilde m$ has the following construction,
\[
\tilde m[X(t),\tau(t)]=\begin{cases}
    \infty, & \text{if $x(t)\in\mathcal T$},\\
    0, & \text{if $x(t)\in\mathcal T'$},
\end{cases}
\]
such that if $x(t)\in \mathcal T$, the therapy fails, and if $x(t)\in \mathcal T'$, the therapy succeeds. We will now outline the explicit forms of the drift and diffusion components in the SDE \eqref{2}. As discussed in \cite{wang2024threshold}, cancer cells can be classified into three categories. Glycolytic cells (GLY) function in the absence of oxygen, producing lactic acid, which can damage nearby healthy tissue. The other two types are aerobic cells, which benefit from enhanced vasculature promoted by the VEGF signaling protein. Among these, VEGF-producing cells (VOP) allocate some of their resources to vascular development, while the remaining aerobic cells do not contribute to this process and can be considered defectors or free-riders, as described in game theory.

We demonstrate our methodology by building upon a model introduced by \cite{carrere2017optimization}, which examines the population sizes of lung cancer cells in vitro. The study focuses on a heterogeneous tumor comprising two types of lung cancer cells: the drug-sensitive (S) ``A549” cells (sensitive to ``Epothilene”) and the drug-resistant (R) ``A549 Epo40” cells  \citep{wang2024threshold}. This model is based on data from experiments conducted by Manon Carrè at the Center for Research in Oncobiology and Oncopharmacology, Aix-Marseille Université. Mutation events were excluded from consideration due to their infrequency at the Epothilene dosages used and the relatively short treatment periods. The competition model, outlined below, was developed based on phenotypic observations, using fluorescent markers to distinguish and track the cells.

Each cell type, when considered independently, follows a logistic growth model with intrinsic growth rates $g_S$ and $g_R$, respectively. The carrying capacity of the Petri dish ($C$) is shared between the two, with resistant cells assumed to occupy $m$ times more space than sensitive cells. Thus, the proportion of space occupied at time $s$ is given by 
\[
\frac{1}{C}\bigg[Z_S(s) + q Z_R(s)\bigg].
\]
When cultured together, it was observed that sensitive cells outgrow resistant cells rapidly, despite their similar intrinsic growth rates \citep{carrere2017optimization}. To represent this competitive advantage, an additional competition term $-\beta Z_S Z_R$ was introduced to model the rate of change of $Z_R(t)$, with the coefficient $\beta$ determined from experimental data. Additionally, it was assumed that resistant cells are completely unaffected by a specific drug, which decreases the sensitive cell population at a rate of $\alpha Z_S(t) \delta(t)$, where $\delta(t)$ represents the rate of drug delivery and $\alpha$ measures the drug’s effectiveness. Therefore, the explicit structure of of the McKean-Vlasov SDE in Equation \eqref{2} is 
\begin{align}\label{4}
  X&=\begin{bmatrix}
      X_1(s)\\X_2(s)
  \end{bmatrix}
  =\begin{bmatrix}
      \frac{Z_S(s)}{Z_S(s)+q Z_R(s)}\\Z_S(s)+ q Z_R(s)
  \end{bmatrix}, \ 
  \bm B(s)=\begin{bmatrix}
      B_S(s)\\ B_R(s)
  \end{bmatrix},\notag\\
  \mu&= \begin{bmatrix}
  \begin{multlined}
  X_1(s)[1-X_1(s)]\biggr\{[1-X_2(s)](\theta_S-\theta_R)-\a\delta(s)+\be C X_1(s)X_2(s)+\tau(s)\\[-4ex]
  +\left[1-X_2(s)\right]^2\left[\sigma_R^2[1-X_1(s)]-\sigma_S\left[\sigma_S X_1(s)+\sigma_R\right]\right]\biggr\}
  \end{multlined}\\[5ex]
  \begin{multlined}
  X_2(s)[1-X_2(s)]\bigg[\theta_SX_1(s)+\theta_R[1-X_1(s)]\bigg]-\a X_1(s)X_2(s)\delta(s)\\[-4ex]
  -\be C X_1(s)[X_2(s)]^2[1-X_1(s)]+\tau(s)
  \end{multlined}
  \end{bmatrix},\notag\\
 \bm\sigma&=\begin{bmatrix}
\left[1-X_2(s)\right]X_1(s)[1-X_1(s)](\sigma_S-\sigma_R)+\tau(s)\\
X_2(s)\left[1-X_2(s)\right]\bigg[\sigma_SX_1(s)+\sigma_R[1-X_1(s)]\bigg]+\tau(s)
 \end{bmatrix}.
  \end{align}
  In the expressions in the system \eqref{4}, $ B(s)$ is a standard two-dimensional Brownian motion for sensitivity (S) and resistannce (R), with volatilities $\sigma_S$ and $\sigma_R$, respectively, $\a$ is drug efficiency, $\be$ is action on sensitive on resistant, q is the size ratio between S and R cells, $(\theta_s,\theta_R)$ are growth rate for sensitive and resistant cells, respectively, and $\delta:\mathbb R\mapsto[0,\delta_{\max}]$ is a time-dependent intensity of S-targetting therapy. Since, our control $u(s)$ is represented by $\delta (s)$, we are interested in minimizing the treatment cost functional
\begin{equation}\label{5}
    \mathcal J\left[X_1(0),X_2(0),\delta\right]=\E\left\{\int_0^t\delta(s)ds+et+\tilde m(X_1(t),X_2(t),\tau(t)\right\},
\end{equation}
where $e>0$ represents the relative importance of stabilization of cancer growth and failure of the treatment. The stochastic Lagrangian \citep{ewald2024adaptation} of this system is
\begin{multline}\label{6}
    \mathcal L[s, X(s),\delta(s),\tau(s),\lambda(s)]=\E\biggr\{\int_0^t h[s, X(s),\delta(s),\tau(s)]ds+\tilde m[X(t),\tau(t)]\\
    +\int_0^t\bigg[X(s)-x_0-\int_0^s\left[\mu[\nu,X(\nu),\delta(\nu),\tau(\nu)]d\nu-\bm\sigma[\nu,X(\nu),\delta(\nu),\tau(\nu)]d B(\nu)\right.\\
    \left.-\bm\sigma_0[\nu,X(\nu),\delta(\nu),\tau(\nu)]d W(\nu)\right]\bigg]d\lambda(s)\biggr\}.
\end{multline}
The term $\lambda(s)$ represents the Lagrange multiplier. To address Equation \eqref{6}, a specialized adaptation of the stochastic Pontryagin principle, such as the Feynman-type path integral method introduced by \cite{pramanik2020optimization} and further enhanced by \cite{pramanik2024optimization}, can be employed to derive an analytical solution for the system \citep{pramanik2024semicooperation}. In scenarios involving high-dimensional state variables and nonlinear dynamics, as seen in Merton-Garman-Hamiltonian SDEs, constructing a Hamilton-Jacobi-Bellman (HJB) equation numerically becomes increasingly challenging \citep{pramanik2023optimal}. The Feynman-type path integral approach addresses these complexities by offering a localized analytical solution \citep{pramanik2021consensus,pramanik2024bayes}. This method begins with formulating a stochastic Lagrangian, represented by Equation \eqref{6}, for every continuous point in the interval $s \in [0, t]$, where $t > 0$. The interval is then divided into $k$ equal subintervals, and a Riemann measure corresponding to the state variable is assigned to each subinterval. By constructing a Euclidean action function, a Schrödinger-like equation is obtained using Wick rotation \citep{djete2022mckean}. Solutions are identified by imposing first-order conditions on the state and control variables. This method shows significant potential in applications such as cancer research \citep{dasgupta2023frequent,hertweck2023clinicopathological,kakkat2023cardiovascular,khan2023myb,vikramdeo2023profiling,khan2024mp60,vikramdeo2024abstract}.

\section*{Discussions:}
The drug delivery rate serves as a vital component in cancer research when treated as a stochastic control, providing a flexible and adaptive framework for optimizing treatment outcomes while reducing adverse effects. Cancer therapies typically rely on administering drugs to either destroy cancer cells or inhibit their proliferation. However, the biological processes underlying cancer progression are intricate, characterized by nonlinear interactions between malignant cells, healthy cells, the immune system, and the tumor microenvironment. These processes are further complicated by stochastic factors such as patient-specific differences, tumor heterogeneity, and environmental influences, making it essential to employ a control mechanism capable of real-time adaptation. Using the drug delivery rate as a stochastic control variable enables researchers and clinicians to modify drug dosages dynamically based on the evolving state of the disease. This approach is particularly valuable in addressing the challenge of treatment-resistant cancer cells, which can arise during therapy due to genetic mutations or adaptations within the tumor microenvironment. By fine-tuning the delivery rate to address such complexities, stochastic models can effectively reduce resistance and extend the efficacy of treatments. Moreover, regulating the drug delivery rate helps to limit toxicity in healthy tissues, a significant drawback of traditional chemotherapy. This is achieved by maintaining drug levels within a therapeutic range that maximizes benefits while minimizing harmful side effects. Advances in computational modeling techniques, such as the HJB equation and path integral approaches \citep{pramanik2024measuring}, have further refined the ability to design and evaluate stochastic drug delivery strategies, offering valuable guidance for optimizing treatment protocols. In the era of precision medicine, these control-based approaches support the development of individualized treatment plans, ultimately improving survival rates and enhancing the quality of life for cancer patients. Consequently, incorporating drug delivery rate as a stochastic control in cancer research addresses critical challenges related to the unpredictable and complex nature of oncological systems.
\bibliographystyle{apalike}
\bibliography{bib}
\end{document}